\documentclass[showpacs,twocolumn,preprintnumbers,amsmath]{revtex4}

\usepackage{graphicx}
\usepackage{dcolumn}
\usepackage{bm}
\usepackage{setspace}
\usepackage{color}
\usepackage{overpic}

\newcommand{\sro}{{$^{88}$}Sr}

\usepackage[english]{babel}

\newlength{\sep}
\newcommand{\ket}{\hspace{\sep}\rangle}

\newcommand{\spc}{\hspace{\sep}|\hspace{\sep}}

\draft

\begin{document}

\title{Engineering the quantum transport of atomic wavefunctions over macroscopic distances}

\author{A. Alberti, V. V. Ivanov, G. M. Tino and G. Ferrari}
\affiliation{Dipartimento di Fisica and LENS - Universit\`a di Firenze, CNR-INFM,\\ INFN - Sezione
di Firenze, via Sansone 1, 50019 Sesto Fiorentino, Italy }
\maketitle

The manipulation of matter waves represents a milestone in
the history of quantum mechanics. The first experimental
validation of matter wave behavior occurred with the observation
of diffraction of matter by crystals\cite{davisson27}, and then
with grating and Young's double-slit interference with electrons,
neutron, atoms and molecules
\cite{joensson61,zeilinger88,carnal91,arndt99}. More recently
matter wave manipulation has become a building block for
quantum devices such as quantum sensors \cite{peters99} and it
plays an essential role in many proposals for implementing quantum
computers \cite{monroe96,monroe02}. In this letter we
demonstrate coherent control of the spatial extent of an
atomic wavefunction by reversibly stretching and shrinking the
wavefunction over a distance of more than one millimeter. The
remarkable experimental simplicity of the scheme could ease
applications in the field of quantum transport and quantum
computing.

Ultracold atomic gases trapped in optical lattices (large and
periodic ensembles of optical microtraps created by interfering
optical laser beams) provide ideal tools for studying
quantum transport in different regimes
\cite{madison98,broawaeys05} and quantum many body systems in
periodic potentials \cite{bloch05,gemelke05,sias07,sias08}. One of
the challenges in this field is to coherently transfer matter
waves between macroscopically separated sites. This would
provide a mechanism to couple distant quantum bits and ultimately
would lead to quantum information processing with cold atoms in
optical lattices \cite{bloch08}. Recently it was demonstrated that
spatially driven lattice potentials in the presence of a linear
potential can induce a coherent delocalization of a matter wave
\cite{ivanov08} when the driving is applied at a frequency
equivalent to the Bloch frequency $\nu_B$, i.e., the linear
potential between adjacent sites expressed in frequency units. The
delocalization occurs at integer multiples of $\nu_B$ because of
the resonant coupling between Wannier-Stark levels within the same
band. The resonances are characterized by a ${\rm sinc}(2 \pi\,
t\Delta\nu)$ spectral profile, where $t$ is the driving time and
$\Delta\nu$ is the detuning of the driving from the resonant
frequency. The ${\rm sinc}$ response here arises from the
influence on the tunneling current of the relative phase $\phi$
between the driving and the site-to-site quantum phase in the
broadened wavefunction. When $\phi$ lies between 0 and $\pi$ the
wavefunction expands, while when it lies between $\pi$ and $2\pi$
the wavefunction shrinks. In particular when $\phi=2\pi$ the
wavefunction returns to the starting point. Such a reversible
behavior is expected provided that the evolution of the
wavefunction is fully coherent.

Any mechanism introducing loss of coherence would in fact lead to
a non-reversible broadening. However, in a decoherence free
regime, it should be possible to engineer the spatial extension of
the wavefunction using the frequency offset and amplitude of the
driving as tuning knobs. Here we experimentally demonstrate this
new technique of matter wave manipulation by showing that coherent
delocalization results in an extended distribution corresponding
to the size of the broadened wavefunction. This is demonstrated by
observing \textit{in-situ} the breathing of the wavefunction under
non resonant driving conditions, and through a self interference
technique based on time-of-flight expansion. Our experimental
findings are supported by a theoretical model with which,
using the basis of Wannier-Stark eigenstates of the static
Hamiltonian, we can determine analytically the spatial
wavefunction under the action of the driving.

\begin{figure}[t] \vspace{-0mm} \begin{center}
\hspace{-0mm}
\includegraphics[width=0.33\textwidth,angle=0]{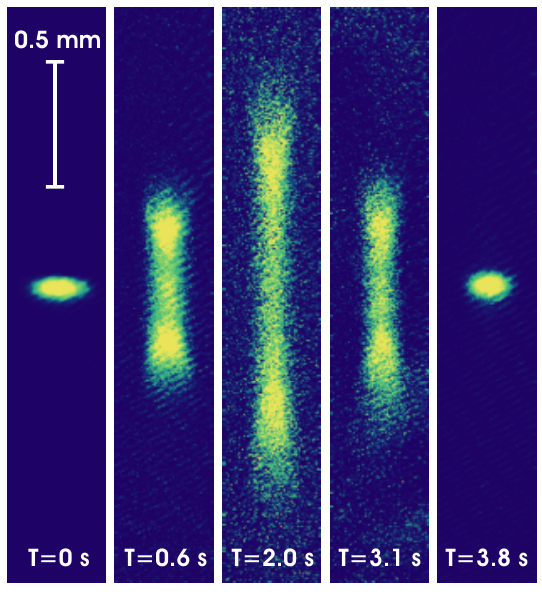}

\vspace{-0mm} \caption{\label{Broadening15mmPicture} Revivals of
the spatial distribution of the atoms in the lattice potential
under strong, non-resonant driving with an amplitude of 10
lattice sites peak-to-peak. In the picture the lattice is
vertically aligned and the driving is active for times
ranging from 0 to 3.8 seconds. The frequency detuning of the
driving $\delta$ is set to about 250 mHz, which results in a
revival time of 3.8 seconds. At T=0 the \textit{in-situ} RMS
spatial width is $31\, \mu$m, while at T=3.8 s it becomes $40\,
\mu$m. The color scale is adapted to each picture so that
visibility is maintained with the varying atomic densities.}
\end{center}
\end{figure}

In order to drive (i.e., modulate) the phase of the lattice
potential, we apply a sinusoidal voltage (with frequency
$\nu_{PZT}$) to the piezo-electric transducer (PZT) that supports
the retro-reflecting mirror of the optical dipole standing-wave.
In a first set of measurements we test the spatial coherence
of the broadened wavefunction by observing the periodic breathing
of the atomic distribution while we drive the PZT with a non-zero
frequency detuning $\Delta\nu=\nu_{\rm PZT}-\nu_B$. Fig.
\ref{Broadening15mmPicture} displays the image of the atomic
distribution under hard driving conditions. The frequency detuning
$\Delta\nu$ is set to about 0.26 Hz such that revival period is
about 4 seconds and the amplitude of the driving is set to
its maximally experimentally accessible value, ten lattice
sites peak-to-peak. The spatial profile is initially
gaussian, corresponding to how the atoms are loaded into the
lattice potential.  The profile then evolves into a more complex
shape as a result of the wavefunction broadening
\cite{thommen04,thommen04bis}, but then at the
revival the distribution returns to its initial profile. Starting
from a size of 31 $\mu$m it reaches an extension larger than 1.5
mm, and then it returns to a size of 40 $\mu$m. In other
words, the distribution increases its size by a factor
larger than 20 and then returns to almost to its initial
value \cite{CoherenceDecay}. Intermediate values of the spatial extent can be obtained
in a reproducible and reversible way by varying the amplitude
or frequency detuning $\Delta\nu$ of the driving voltage.

\begin{figure}[t] \vspace{-0mm} \begin{center}
\hspace{-1mm}
\includegraphics[width=0.48\textwidth,angle=0]{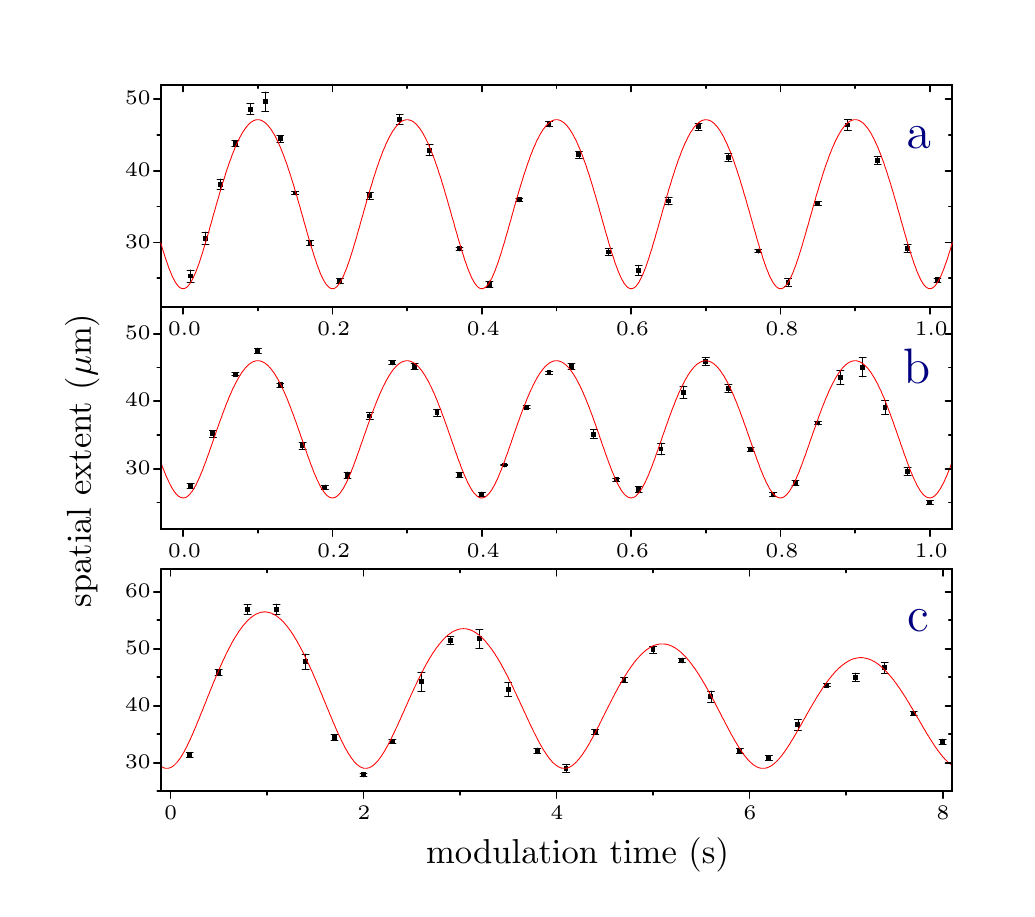} 

\vspace{-0mm} \caption{\label{SmallSpatialBroadeningGraph}
Dynamics and revivals of the atomic extent under non-resonant
driving of the lattice phase. For short times the minimum
size corresponds to the size of the non-driven atomic cloud. a)
detuning $\delta=\nu_{\rm PZT}-\nu_B=+5$ Hz, b) $\delta=-5$ Hz, c)
$\delta=-0.5$ Hz. The driving amplitude is set to about 1 (a and
b) and 0.2 (c) lattice sites peak-to-peak. From the data in
c) we infer a $e^{-1}$ damping time of 28 s. In each graph the
fitted line gives, within the error bars, a periodicity equal to
$\delta^{-1}$.}
\end{center}
\end{figure}

In Fig. \ref{SmallSpatialBroadeningGraph} we show the time evolution of the spatial extent for different values of
frequency detuning $\Delta\nu$. In Fig. \ref{SmallSpatialBroadeningGraph}a and b, with $\Delta\nu$ equal to $+5$ Hz
and $-5$ Hz respectively, the breathing shows a revival with the expected period of 5 Hz, and a constant visibility
on a one second timescale, regardless of the sign of the frequency detuning. Fig.
\ref{SmallSpatialBroadeningGraph}c is similar to Fig. \ref{SmallSpatialBroadeningGraph}a but with $\Delta\nu=0.5$
Hz. Again the atomic distribution shows a breathing at a frequency equal to $\Delta\nu$, and from the
reduction of the oscillation amplitude over time we can infer a $e^{-1}$ damping time of 28 seconds. The results
presented in Fig. \ref{Broadening15mmPicture} and \ref{SmallSpatialBroadeningGraph} can not be explained
classically and show a quantitative agreement with the analytic expression of the wavefunction expected for the
driven potential \cite{thommen04,thommen04bis,CoherenceDecay}. This implies that we manipulate, and directly
observe, the spatial wavefunction on a length scale larger than 1 mm.

\begin{figure}[t] \vspace{-0mm} \begin{center}
\hspace{-0mm}
\includegraphics[width=0.47\textwidth,angle=0]{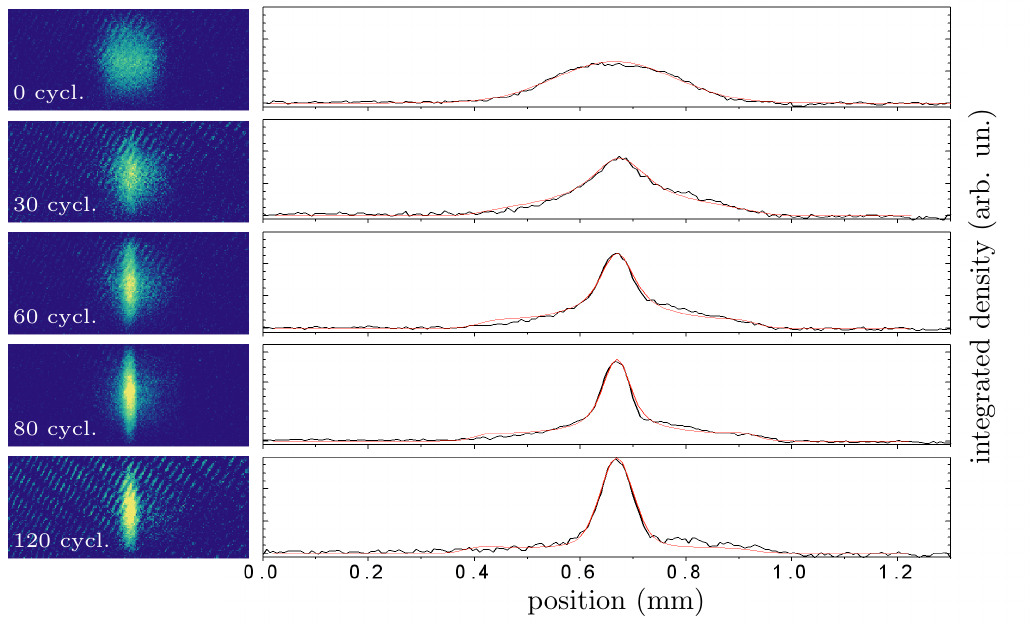}

\vspace{-0mm} \caption{\label{CycleNumber} Effect of the driving
on the ballistic expansion of the sample. The time-of-flight is
fixed to 15 ms and a resonant driving is applied during a fixed
amount of cycles. Left side: 2D density profile. In the pictures
the vertical gravity and the lattice are horizontally oriented.
Right side: the atomic density integrated along the
transverse direction of the lattice. The red line is the
distribution expected from the ballistic expansion of the
analytical expression of the broadened wavefunction
\cite{SimulationTOF}. Both the colour scale on the 2D distribution
and the vertical scale on the integrated density profiles are the
same for the various driving conditions.}
\end{center}
\end{figure}

To study the coherence properties of the modified wavefunction we measure the interference of the wavefunction with
itself. To this end, the wavefunction is expanded in time-of-flight (TOF) after an adiabatic release from the
lattice potential. We record the evolution for the interval between 0 to 25 ms under various driving conditions.
First we consider the momentum distribution under resonant driving(i.e., $\Delta\nu=0$). An integer number of
sinusoidal cycles, up to 120 (equivalent to 210 ms when $\nu_B=574.14$ Hz), is applied to the PZT such that the
wavefunction broadens proportionally with time. The momentum distribution then is probed in TOF 300 ms after the
beginning of the driving, and after switching off the lattice potential adiabatically on a 20 $\mu$s timescale.
Fig. \ref{CycleNumber} shows the changes in the expansion after 15 ms of TOF for various broadening conditions. In
the absence of a driving voltage, the thermal sample expands as expected yielding a gaussian profile. When we drive
the system, we clearly observe the appearance of a non-Gaussian distribution. Here we can distinguish two
components: the first is directly related to the expansion in absence of the modulation, while the second one has a
spatial extent and relative weight related to the broadening. This results from the interference among the
probability amplitude originating from the different portions of the wavefunction as directly checked by a
simulation of the the ballistic expansion of the broadened wavefunction. As expected, when the free expansion due
to the momentum dispersion is of the order of the size of the broadened wavefunction, the interference pattern reaches maximum visibility.
\begin{figure}[t] \vspace{-0mm} \begin{center}
\hspace{-1mm}
\includegraphics[width=0.4\textwidth,angle=0]{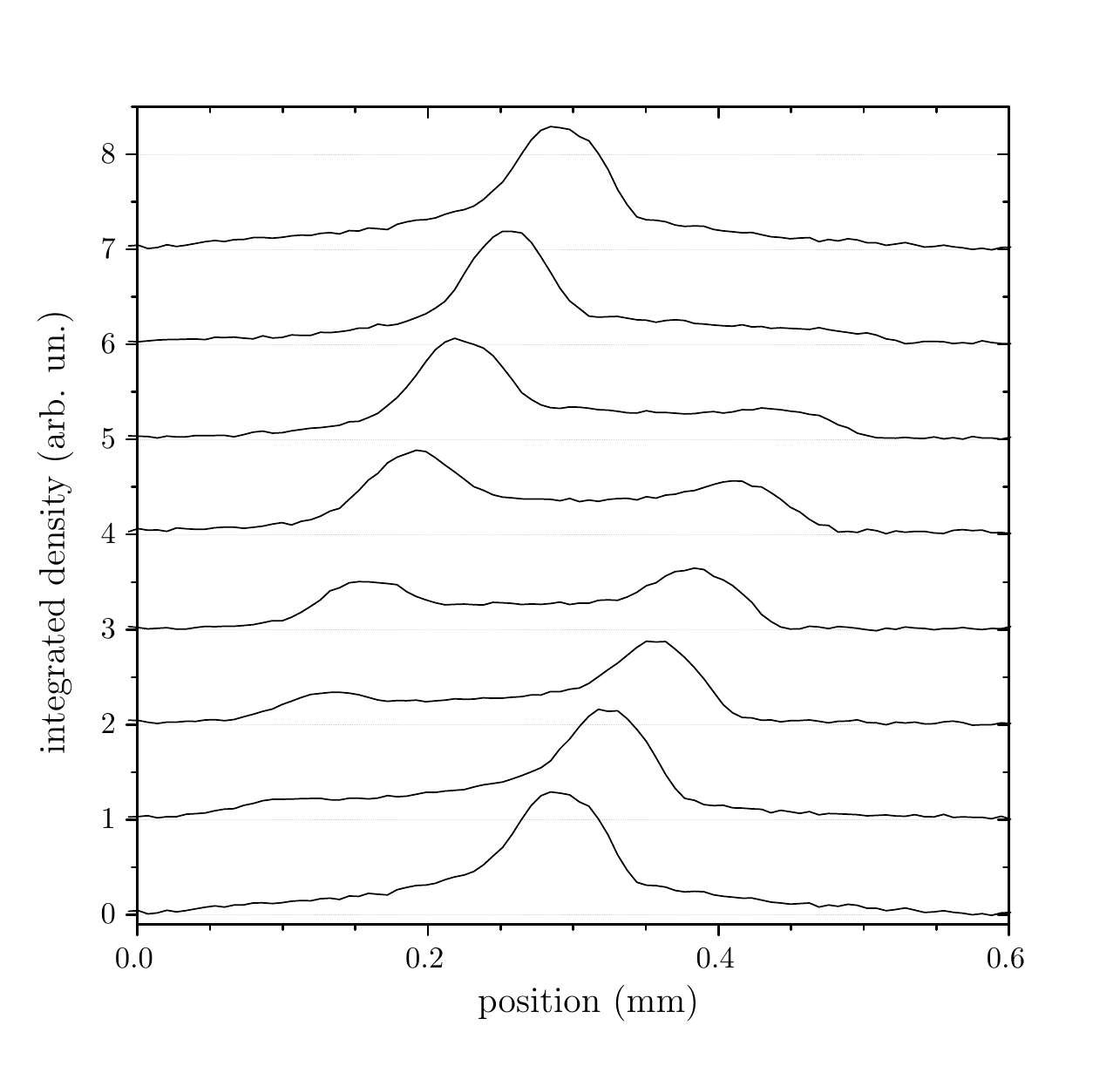}  

\vspace{-0mm} \caption{\label{BlochOsc} Temporal evolution of the
broadened wavefunction phase in the absence of a driving field. In
a static potential the combined effect of the lattice and gravity
induces a linear increase of the site-to-site phase in the
broadened wavefunction. This is observable in the ballistic
expansion signal as a periodic evolution of the interference peak
for a fixed time-of-flight duration of 15 ms. From top to bottom
the trapping time is linearly increased over one Bloch period
(about 1.75 ms).}
\end{center}
\end{figure}

There is an additional effect that, in principle, can complicate
the observed TOF profiles. For a static lattice (i.e., one without
a driving force applied to the PZT) the position of the
interference peak exhibits periodic dynamics at the Bloch
frequency $\nu_{B}$. This result, shown in Fig. \ref{BlochOsc},
reflects the time evolution of the site-to-site phase differences
in the wavefunction and is confirmed by simulation of the
wavefunction expansion. This phenomenon is in direct analogy with
Bloch oscillations where the atoms, subject simultaneously to a
constant force (i.e., gravity in this case) and the lattice
potential, sweep periodically the first Brillouin zone
\cite{dahan96}.  In the TOF images this is observed as an increase
of the momentum of the interference peack linearly in time, until
it reaches the lattice photon recoil and changes sign. At this
point the interference peak is split in two parts. In order to
simplify the treatment and neglect this additional effect, the
pictures of Fig. \ref{CycleNumber} and the subsequent analysis are
made by starting the TOF at point where the interference
peak has maximum amplitude or, equivalently, it is centered in the
first band before the release from the lattice.

We have confirmed both the appearance of the interference
pattern and its Bloch like dynamics by simulating the evolution in
the static lattice and the free expansion of the analytic form of
the broadened wavefunction \cite{thommen04}:

\begin{multline}
\spc n(t)\ket=\hspace{-4pt}\sum_{n'=-\infty}^{+\infty}\hspace{-5pt}e^{-i\hspace{0.6pt}n'\hspace{-0.0pt}2\pi\hspace{+0.5pt}\nu_B t}e^{i\hspace{0.6pt}\pi\hspace{0.6pt}(n-n')\hspace{0.6pt}\Delta \nu \, t}\hspace{3pt}\times\\
\times\hspace{3pt}J_{n-n'}\Big(\sin(\pi \Delta \nu\, t)\frac{2 \Omega}{\pi\Delta\nu}\Big)\,\spc n'\ket
\label{FinalWavefunction}
\end{multline}
where $\spc n'\ket$ are the Wannier-Stark eigentates of the static Hamiltonian (with the gravity potential and the
non-modulated lattice) labelled by the index $n'$ of their position expressed in lattice units, $\spc n(t)\ket$ is
wavefunction of the state $\spc n \ket$ subject to the driving during a time $t$, $\Omega$ is a coefficient
accounting for the tunneling rate among the lattice barriers, and $J_n$ are the Bessel functions of the first kind.
The hypothesis that the initial state corresponds to the Wannier-Stark eigenstate $\spc n\ket$ is justified by the
fact that the trapped atoms are derived from a thermal sample at a temperature higher than the recoil temperature
and they uniformly fill the first band (see the Methods section). The thermal de Broglie wavelength is shorter than
the lattice period, so when we load the atoms into the lattice potential, the coherence properties among adjacent
Wannier-Stark eigenstates can be neglected. In Fig. \ref{CycleNumber} the density profiles we derive from the
analytical model are in remarkable agreement with the experimental data. It is also worth noting that the
Bloch-like dynamics observed in Fig. \ref{BlochOsc} results only from the long range coherence of the
broadened wavefunction. In the absence of driving, on the other hand, the shape of the distribution after ballistic
expansion does not show any periodicity.

\begin{figure}[t] \vspace{-0mm} \begin{center}
\hspace{-1mm}
\includegraphics[width=0.48\textwidth,angle=0]{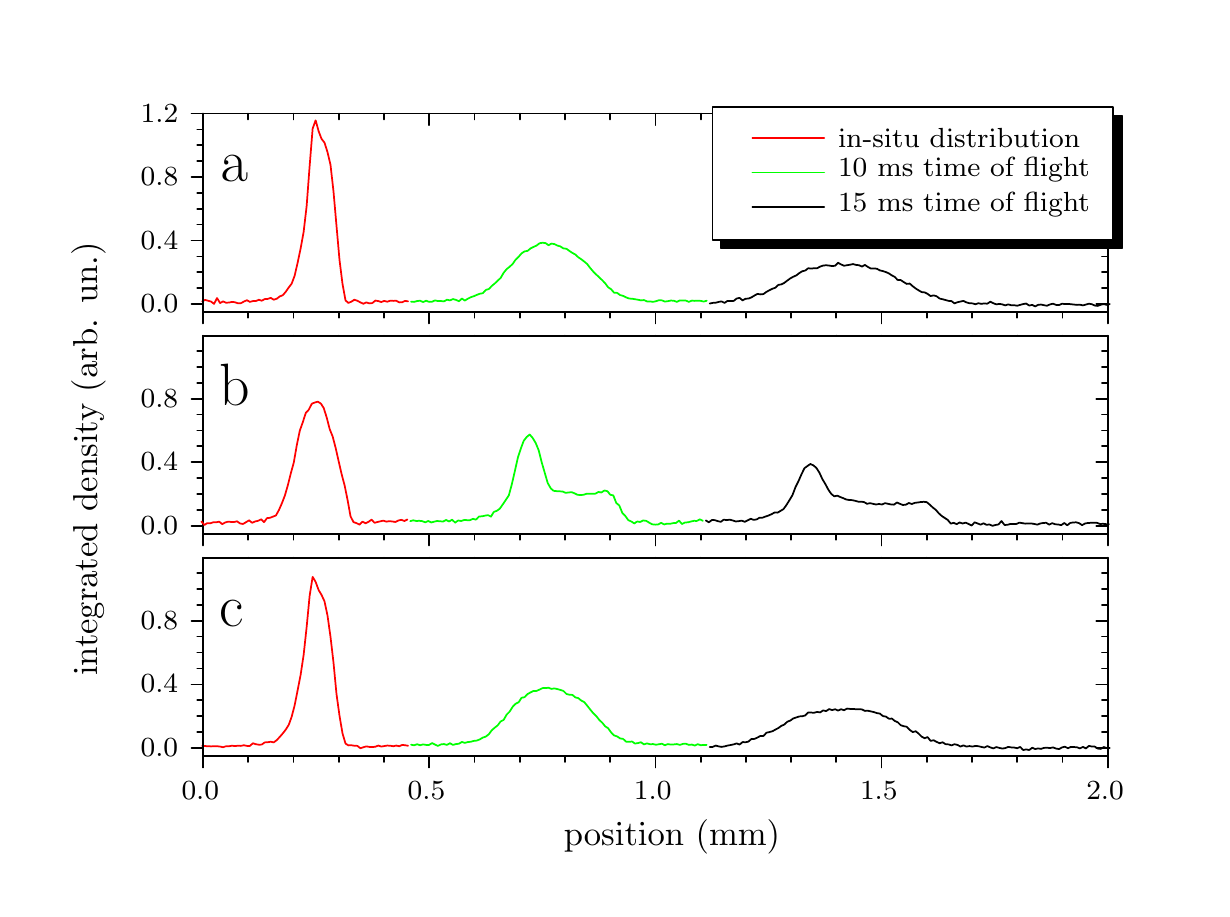}   

\vspace{-0mm} \caption{\label{MomentumRevival} Ballistic expansion
under non-resonant driving. The periodic dynamics of the spatial
extent have their counterpart in the ballistic expansion
signal. From left to right: {\it in-situ} distribution, 10 ms and
15 ms time-of-flight. A: no driving. B: maximum expansion. C:
first spatial revival. In B the driving conditions are the
same as those in Fig. \ref{CycleNumber} for 80 cycles, except
here we use a 2.28 Hz frequency detuning of the driving field
with respect to the Bloch frequency. The horizontal displacement
corresponds to the free fall in the earth gravity field.}
\end{center}
\end{figure}

As an additional check of the agreement between the
experiment and theory we verified that the modified distributions
in the time-of-flight profiles do not result from a simple
selection of a given momentum class. For instance, for the
conditions of Fig. \ref{CycleNumber} the overall atom number does
not change with the number of applied cycles; the  atomic
distribution in this case shifts to the central peak owing to the
constructive interference between probability amplitudes from
symmetric regions of the wavefunction. In addition, we verified
that the wavefunction also exhibits revivals in the interference
pattern for the TOF distribution under non-resonant driving. In
Fig. \ref{MomentumRevival} we compare the TOF profiles taken for
different expansion times under three different conditions:
without driving, at maximum expansion, and at the spatial revival.
The driving conditions for the last condition are similar to those
of Fig. \ref{CycleNumber} except that the frequency detuning
$\delta$ is set to about 2 Hz and the driving lasts longer in
order to reach a maximum broadening as in Fig. \ref{CycleNumber}
with 80 cycles. The TOF expansion at maximum broadening is
equivalent to that obtained under resonant driving except that now
the distribution is deliberately asymmetric because of the
different Bloch phase at the release (same as in fig.
\ref{BlochOsc} case 5). On the other hand, the TOF expansion at
the time when the distribution shrinks to the initial size, is the
same as that in absence of driving.

In summary, we experimentally demonstrates a new method to
manipulate the spatial wavefunction of cold strontium atoms over
distances of the order of a millimeter. This manipulation
preserves the quantum coherence and we find that the process is
fully deterministic, reversible, and in quantitative agreement
with an analytical model. The controlled stretching and shrinking
of the wavefunctions may be applied for implementing new classes
of quantum logic gates for q-bits such as neutral atoms in an
optical lattice \cite{hayes07,creffield07}. It is worth mentioning
that neutral atom optical clocks can operate with atoms in
conditions similar to those presented in this work
\cite{takamoto05,ludlow08}. Finally, with stretched wavefunctions
we observe a striking dependence of the distribution in ballistic
expansion due to site-to-site phase gradients induced by external
forces, regardless of the initial temperature of the sample. This
approach may find applications in the realization of precise force
sensors based on atoms otherwise difficult to cool.

\section{Methods}

\subsection*{Experimental setup and procedure}
The experimental setup was described previously in
\cite{ferrari06bis}. About $10^6$ \sro\ atoms are laser cooled to
a temperature of 1 $\mu$K and subsequently are loaded into a
vertical lattice potential produed by the dipole force of a
$\lambda_L=$ 532 nm laser field. The lattice is formed by
retro-reflecting 532 nm laser light with a mirror mounted on top
of a piezo-electric transducer (PZT). The phase of the optical
lattice, defined to be zero at the surface of the mirror, is
modulated by applying a time-dependent voltage to the PZT with a
maximum excursion of 10 lattice sites peak-to-peak. The depth of
the lattice potential is typically 8 recoil energies $E_R$ ($E_R =
h^2/2 m\lambda_L^2 = k_B \times 381$ nK) along the optical axis of
the trapping beam and, in the radial direction, it decreases
exponentially due to the gaussian spatial profile. Since initially
the atoms are hot with respect to the recoil temperature, when
they are loaded into the lattice potential they occupy almost
uniformly the first band and, with decreasing weight, the higher
bands. Taking into account the full three dimensional lattice
potential and including the gravity force, all the bands except
the first one have a lifetime shorter than 100 ms. We finish the
atom sample preparation sequence by taking advantage of this
property to select atoms in the first band, simply by holding the
atoms in the static potential for a time of 100 ms. The atomic
distribution is measured by absorption imaging of a resonant laser
beam detected on a CCD camera.

\begin{description}
 \item[Acknowledgements] We would like to thank M. Artoni, N. Poli, C. W. Oates  and M. L. Chiofalo for stimulating discussions, M. Schioppo for his contribution in the early stage of the experiment, and R. Ballerini, M. De Pas, M. Giuntini, A. Hajeb, and A. Montori for technical assistance. This work was supported by LENS, INFN, EU (under Contract No. RII3-CT-2003 506350 and the FINAQS project), ASI and Ente CRF.
  \item[Competing Interests] The authors declare that they have no competing financial interests.
 \item[Correspondence] Correspondence and requests for materials
should be addressed to G.F.~(email: ferrari@lens.unifi.it).
 \end{description}

\end{document}